# Deep learning automated quantification of lung disease in pulmonary hypertension on CT pulmonary angiography: A preliminary clinical study with external validation

**Authors**: Michael J. Sharkey*[†,1,2], Krit Dwivedi*[1], Samer Alabed[1], Andrew J. Swift[1,3,4]

* M.S. and K.D. contributed equally to this work.
† Corresponding author Michael.Sharkey3@nhs.net
[1.] Department of Infection, Immunity and Cardiovascular Disease, University of Sheffield, Sheffield S10 2RX, UK
[2.] 3DLab, Radiology Department, Sheffield Teaching Hospitals NHS Foundation Trust, Sheffield S10 2JF, UK
[3.] INSIGNEO, Institute for In Silico Medicine, University of Sheffield, Sheffield S1 3JD, UK
[4.] NIHR Sheffield Biomedical Research Centre, University of Sheffield, S10 2JF, UK

# Abstract

**Purpose**

Lung disease assessment in precapillary pulmonary hypertension (PH) is essential for appropriate patient management. This study aims to develop an artificial intelligence (AI) deep learning model for lung texture classification in CT Pulmonary Angiography (CTPA), and evaluate its correlation with clinical assessment methods.

**Materials and Methods**

In this retrospective study with external validation, 122 patients with pre-capillary PH were used to train (n=83), validate (n=17) and test (n=10 internal test, n=12 external test) a patch based DenseNet-121 classification model. 'Normal', 'Ground glass', 'Ground glass with reticulation', 'Honeycombing', and 'Emphysema' were classified as per the Fleishner Society glossary of terms. Ground truth classes were segmented by two radiologists with patches extracted from the labelled regions. Proportion of lung volume for each texture was calculated by classifying patches throughout the entire lung volume to generate a coarse texture classification mapping throughout the lung parenchyma. AI output was assessed against diffusing capacity of carbon monoxide (DLCO) and specialist radiologist reported disease severity.

**Results**: Micro-average AUCs for the validation, internal test, and external test were 0.92, 0.95, and 0.94, respectively. The model had consistent performance across parenchymal textures, demonstrated strong correlation with diffusing capacity of carbon monoxide (DLCO), and showed good correspondence with disease severity reported by specialist radiologists.

**Conclusion:**

The classification model demonstrates excellent performance on external validation. The clinical utility of its output has been demonstrated. This objective, repeatable measure of disease severity can aid in patient management in adjunct to radiological reporting.

# 1. Introduction

Pulmonary hypertension (PH) is a progressive, heterogenous, incurable condition with significant morbidity and mortality. Computational Tomography (CT) imaging plays an integral role in PH management and is recommended by the latest joint European Respiratory Society/European Society of Cardiology PH guidelines and Pulmonary Vascular Research Institute (PVRI) statement on imaging in pulmonary hypertension (1,2). There is growing evidence for better quantifying and characterising lung disease in patients with pre-capillary PH (3,4).

Existing work in the literature on classification of lung parenchymal texture patterns has been performed in non-contrast imaging, on diseases such as emphysema, COPD or interstitial lung disease(5–9). Within PH, CT Pulmonary Angiography (CTPA), which involves administration of intravenous contrast, is routinely performed as it allows for simultaneous assessment of the vascular structures. There are significant differences in parenchymal appearances between non-contrast and CTPA imaging, secondary to the lung tissue uptake of contrast media. On review, we found only one preliminary study assessing lung parenchymal patterns in contrast enhanced Chest CT, finding significant difference in mean lung density in patients with pulmonary embolism(10).

Prior work has demonstrated the capability of a patch based approach to classify lung texture in CTPA imaging (11). The aim of this study was to develop a deep learning lung texture classification model optimised with a greater number of hyperparameters, determine generalisability in external data and clinically evaluate how the models quantitative outputs correlate with established clinical methods of assessing lung disease severity.

# 2. Methods

**Study cohort**

Patients from the ASPIRE (Assessing the Severity of Pulmonary Hypertension In a Pulmonary Hypertension REferral Centre) registry with a diagnosis of precapillary pulmonary hypertension were selected (12). The registry includes multimodality imaging and clinical data from patients referred to a specialist tertiary referral centre. Inclusion criteria for imaging were thin slice (<1.25mm), CTPA phase scans with intravenous contrast. Patients with incomplete scans without the entirety of the lung parenchymal or with severe imaging artefacts were excluded.

**Datasets**

The study cohort (see figure 1) was divided 68%/14%/8%/10% into training, validation and internal and external test sets. 'Training' is defined as data on which the model is initially fit. 'Validation' is unbiased testing of each trained model and used to evaluate the hyperparameter tuning. The 'test' datasets are an unseen unbiased dataset on which the final model is evaluated a single time.

## Classification model development

A patch based method to achieve classification of the lung parenchyma on CTPA imaging has been developed. The model classifies between five classes, as defined by the Fleishner Society glossary of terms - 'normal lung', Pure 'ground glass' (GG), 'ground glass with reticulation' (GGR), 'emphysema' and 'honeycombing' (13). The approach is visualised in Figure 2.

### Manual expert labelling

Two radiologists, with 12 and 2 years of specialist experience in cardiothoracic imaging labelled regions from each CTPA scan for the chosen five parenchymal patterns. Care was taken to ensure a range of severity of the underlying disease pattern (for example mild, moderate and severe emphysema) was included. Regions were segmented using MIM Software (MIM Software, Cleveland, Ohio, US).

### Patch creation and image atlas

The manually segmented texture regions were compiled into a texture atlas, patches were then selected randomly from the atlas based on the centre pixel of the patch being labelled as a specific texture. Patches could be extracted from the texture atlas with a variety of model hyperparameters:
- Patch size - square or cubic patches of N pixels. N was varied from 8 to 64.
- Patch dimensionality - 2D, 2.5D or 3D, the 2.5D patch consisted of 2D patches aligned to the orthogonal imaging axes - axial, coronal and sagittal, these are stacked into a NxNx3 stack for training.
- Selection radius - this parameter determines how close patches can be to each other and is used to prevent closely correlated patches being included in the training and testing data. The selection radius was varied from 1 - 10mm and created a sphere in which no other patch can be selected.
- Minimum fill factor - this parameter corresponds to the proportion of a selected patch which has been manually labelled as a specific texture. Minimum fill factor was varied between 0.5 and 1 for the training patches, validation patches had a minimum fill factor of 0.5.
- Number of patches - this parameter corresponds to the volume of lung data that the model sees during training. The training data always had balanced numbers of patches from each texture, therefore the texture with minimum manually segmented volume was the limiting factor for the number of training patches for each class.

**Deep learning classification model**
A DenseNet 121 model (14) was used to train a 5 class classification network. The model has a single input (a single patch) with 5 output classes, one for each texture. Patches were normalised to give an intensity between 0 and 1, and random rotations (range -15-15º), zoom (range 0.9-1.1) and flip along axis were applied with a probability of 0.5. Variable batch sizes (200-8000) were used for each experiment depending on the patch size. For 2D patches a 2D 4 block densenet with 6, 12, 24 and 16 connections per block was trained, for 2.5D patches a 3D network with a single block with 24 connections was used and for 3D patches a 3D network with varying block configuration was used: 2 blocks with 6 and 24 connections for patch sizes between 5 and 12, a 3 block with 6, 24 and 16 connections for patch sizes between 13 and 32 and a 4 block densenet with 6, 12, 24 and 16 connections for all other patch sizes. In all networks there were 64 initial filters and a growth rate of 32 filters per block. To prevent overfitting and to provide a fair comparison between the different experiments, training was terminated when accuracy scores hadn't improved in 60 epochs (patience).

These hyperparameters were tuned:
- Patch dimensionality of 2D, 2.5D and 3D
- Patch sizes of 8, 16, 24, 32, 48, and 64 pixels square or cubed depending on the dimensionality of the patch
- Selection radius of 1, 3, 5 and 10mm
- Minimum fill factor of 0.5, 0.625, 0.75, 0.875 and 1
- Number of patches per class of 300, 1000, 3000 and 10000

Optimal hyperparameters were determined by the combination giving the highest AUC in the validation cohort using a five-fold cross-validation hyperparameter search.
A final model was trained using the optimum hyperparameters and the entire training library. This model was evaluated in the internal validation, internal test and external test datasets.

**Reconstruction of classified patches to quantify entire lung parenchyma**
The reconstruction workflow segments the lung parenchyma using a pre-trained nnUNet. A sliding window with stride of 8x8x1 pixels is used to extract overlapping patches in a grid throughout the lung parenchyma with a spacing of 8 pixels in x and y directions and every slice in the z direction. Each extracted patch is classified using the trained DenseNet model and the 8x8x1 pixels centred on the patch origin are updated to the predicted class in order to generate a coarsely labelled volume. This stride length was found to give visually appealing and interpretable results however this hyperparameter was not optimised in this work.

## Clinical validation

The classification reconstruction provides total lung volume and volume for each of the five parenchymal classes. Percentage of lung involvement for each class was computed. This was clinically validated by correlation against diffusing capacity of carbon monoxide (DLCO), and graded severity on routine clinical reporting. CT scans were evaluated by specialists radiologists for the presence of fibrotic or emphysematous changes, graded as absent, mild, moderate, or severe, as previously described (4,15).

## Statistics

All analysis is performed retrospectively at the patient level with one single corresponding incident CTPA. Analysis was performed with R version 4.1.2. Categorical data are presented as number and percentage, continuous data as median and IQR. To compare between datasets, two-sample Welch t-tests or Wilcoxon rank sum tests were used for continuous data. Categorical data were compared by Pearson's $\chi^2$ test or by Fisher's exact test. Spearman's correlations with DLCO and association with radiologically reported disease severity was performed on a full cohort level. Models were evaluated using the multiclass receiver operator characteristic area under the curve (AUC).

# 3. Results

From 5643 patients in the ASPIRE registry, 521 patients between February 2001 and January 2019 met the inclusion criteria, 122 patients were randomly selected to form the study cohort. 12 patients with imaging performed outside ASPIRE from 9 different centres were set aside as holdout external testing. The remaining 110 patients were randomly split between 83 for training, 17 for validation and 10 for holdout internal testing. Patient clinical characteristics and scanner data is described in Table 1. There are no statistically significant differences in clinical characteristics between the different cohorts. The external cohort differed from the internal cohorts in the range of CT scanner manufacturers consisting of predominantly Siemens and Canon compared to predominantly General Electric (GE) in the internal cases.

**Technical results**

A hyperparameter search found the optimum parameters were 2.5D patches, pixel size of 64, selection radius of 3mm, minimum fill factor of 0.625 and the maximum number of patches. A final model with these hyperparameters was trained in the full patch library consisting of 48093 patches per class. ROC curves and AUCs for the final model are presented in Figure 3, and Table 2 respectively for the evaluation in the internal validation cohort of 10392 patches per class, the internal test cohort of 10891 patches per class, and the external test cohort of 8348 patches per class.

Micro-average AUCs for the internal validation, internal test and external test were 0.92, 0.95 and 0.94 respectively. GGR consistently had the worst performance with AUCs of 0.87, 0.88 and 0.87 in the validation and internal and external test cohorts respectively. Emphysema had the largest differences in performance with AUC varying from 0.87 in the validation cohort to 0.99 and 0.96 in the internal and external test cohorts respectively. Normal, GG and Honeycombing had similar performance with AUCs of greater than 0.9 in all cohorts.

**Clinical validation**

In the full cohort, there was strong positive correlation (R = 0.63) between AI quantified percentage of normal lung parenchyma and DLCO. Honeycombing, GGR, and emphysema had moderate negative correlation (R = -0.49, -0.33 and -0.42 respectively). GG had very weak negative correlation (R= -0.13).

AI quantified features corresponded well with the disease severity reported by experienced specialist radiologists. 'None', 'mild', 'moderate' and 'severe' emphysema radiologically corresponded to 1(0,9) %, 4(2,14)%, 31(21,51)%, and 69(33,77)% AI quantified emphysema; there was a significant (p<0.001) difference between groups.

'None', 'mild', 'moderate' and 'severe' fibrosis radiologically correspond to 3%, 15%, 24% and 44% AI quantified fibrosis, 2%, 8%, 8% and 12% GG, 1%, 11%, 18% and 30% GGR, and 0.8%, 2.0%, 6.8% and 12.5% honeycombing respectively. There was a significant (p<0.001) difference between groups for all features.

## 4. Discussion

This study describes a lung parenchyma classification model in pulmonary hypertension with excellent technical performance and external validation. The clinical utility of the model output has been demonstrated by correlation with DLCO and associated with expert radiologically reported disease severity. To the best of our knowledge, this is the first study to demonstrate this in CTPA imaging or PH.

A major challenge in building supervised machine learning models for medical imaging is the expense and time commitment required for high quality labelled training data (16,17). Developing a classification algorithm requires subspecialty radiologists to label numerous examples of each class across a large heterogeneous patient cohort. Building such a dataset with expert labels is challenging in a rare condition such as PH. As each CTPA contains over 500 slices, semantically labelling each area of every slice of every scan for each patient is impractical and unachievable. The patch based approach in this study allows for creation of a texture 'atlas' from labelled regions, and does not require exhaustive labelling of all voxels for each slice. This atlas can then subsequently be used to train an existing deep-learning classification architecture; DenseNet 121 in this study (14).

The technical performance achieved by our model (average ROC-AUC 0.94 in external test) is similar to other publications in this domain in non-contrast imaging (5–7). The highest AUC (0.99) was achieved by Bermejo-Peláez et al using an ensemble approach which combined 10 different classification networks for a 8 class parenchyma classification task (5) specific external hold out testing was not performed. Technical performance is not directly comparable, due to the differences in the dataset structure and number of classes. Our model demonstrates good generalisability to other scanner manufacturers and acquisition protocols with similar performance in the internal and external test sets.

Established methods of assessing lung disease severity in PH include DLCO on pulmonary function testing and radiological assessment of CTPA imaging (1,2,18,19). Both methods have respective limitations of significant variability and low reproducibility(20–25). The promise of AI quantified imaging metrics is to provide a new, adjunct and repeatable means of quantifying severity (15). In this study, we demonstrate moderate to strong correlation of our AI quantified disease percentage with DLCO. We further demonstrate good agreement between AI and specialist radiologist quantified emphysema and fibrosis. This adds confidence in their use clinically, and further studies in this area are warranted.

## 5. Limitations

The findings should be interpreted in context with several limitations. Although the clinical results are validated across multiple scanners and centres, training and testing was performed on imaging from a single tertiary centre with imaging from modern, high quality thin-slice scanners. External validation using larger and more diverse datasets from multiple institutions is necessary to confirm the generalizability of our findings. All classification models have inherent, difficult to quantify, biases skewed by their development training set. The classification model will learn from the preferences, biases, and opinions of the radiologists that labelled the training data. Other models which differentiate between more parenchymal classes exist. However, the choice of the five parenchymal classes chosen in this study was informed by prior work on the impact of lung disease in precapillary PH (3,4,15). In the future, we aim to involve more radiologists and of differing expertise and multicentre data to create a more representative model.

## 6. Conclusion

In conclusion, this study presents a deep-learning patch based approach to lung parenchyma classification on CTPA imaging in PH. The hyperparameters of this approach have been tuned, and the model demonstrates excellent performance on external testing. The AI output quantifying the percentage of lung involvement corresponds with the current established methods of lung disease severity assessment. This approach can be utilised within pulmonary hypertension to study the impact of these parenchymal patterns on phenotyping, diagnosis and prognosis.

| Characteristic | Full Cohort N = 122[1] | training, N = 83[1] | validation, N = 17[1] | internal test, N = 10[1] | external test, N = 12[1] | p-value[2] |
|---|---|---|---|---|---|---|
| Age at diagnosis, years | 68 (60, 75) | 68 (60, 75) | 66 (61, 75) | 67 (54, 78) | 69 (65, 73) | >0.9 |
| Sex, female | 58 (49%) | 37 (46%) | 10 (59%) | 5 (50%) | 6 (50%) | 0.8 |
| Body Mass Index | 27.2 (23.8, 31.8) | 27.4 (23.8, 30.8) | 27.6 (25.0, 32.6) | 25.7 (23.6, 27.1) | 25.9 (21.9, 33.3) | 0.4 |
| WHO Function class | | | | | | 0.9 |
| 2 | 10 (8.5%) | 7 (8.8%) | 1 (6.2%) | 1 (11%) | 1 (8.3%) | |
| 3 | 58 (50%) | 42 (52%) | 8 (50%) | 4 (44%) | 4 (33%) | |
| 4 | 49 (42%) | 31 (39%) | 7 (44%) | 4 (44%) | 7 (58%) | |
| FVC, percent predicted | 78 (65, 103) | 80 (65, 104) | 78 (67, 105) | 67 (58, 110) | 73 (63, 74) | 0.7 |
| FEV1, percent predicted | 69 (53, 85) | 70 (56, 85) | 67 (48, 89) | 73 (51, 76) | 64 (51, 80) | >0.9 |
| FEV1 / FVC ratio | 73 (59, 81) | 72 (58, 81) | 66 (56, 78) | 73 (63, 79) | 77 (67, 84) | 0.5 |
| DLCO, percent predicted | 26 (18, 37) | 27 (18, 42) | 23 (19, 27) | 25 (22, 27) | 27 (17, 31) | 0.3 |
| mPAP, mm Hg | 46 (35, 54) | 45 (34, 56) | 47 (40, 56) | 39 (28, 48) | 46 (41, 50) | 0.3 |
| CT scanner manufacturer | | | | | | **<0.001** |
| GE | 107 (90%) | 80 (100%) | 17 (100%) | 9 (90%) | 1 (8.3%) | |
| Siemens | 6 (5.0%) | 0 (0%) | 0 (0%) | 0 (0%) | 6 (50%) | |
| Canon | 6 (5.0%) | 0 (0%) | 0 (0%) | 1 (10%) | 5 (42%) | |
| Normal lung | 43 (20, 63) | 46 (30, 70) | 26 (19, 62) | 25 (15, 42) | 34 (17, 53) | 0.058 |
| Ground glass (GG) | 5 (1, 13) | 4 (1, 10) | 5 (2, 20) | 10 (2, 19) | 7 (4, 11) | 0.3 |
| Ground glass with reticulation (GGR) | 6 (1, 24) | 4 (0, 17) | 14 (2, 26) | 23 (3, 36) | 7 (2, 30) | 0.11 |
| Honeycombing | 2 (1, 7) | 2 (1, 6) | 2 (1, 13) | 4 (2, 10) | 5 (2, 9) | 0.2 |
| Emphysema | 10 (1, 33) | 15 (1, 32) | 8 (2, 26) | 13 (0, 46) | 1 (0, 12) | 0.4 |
| Fibrosis | 11 (2, 32) | 6 (1, 26) | 18 (3, 42) | 30 (8, 41) | 10 (6, 40) | 0.10 |

[1] Median (IQR); n (%)
[2] Kruskal-Wallis rank sum test; Fisher's exact test

**Table 1: Patient characteristics for the full cohort, training, validation, internal and external test datasets.** Categorical data are shown as n and (%) of the respective population. Continuous data are depicted as median [Q1, Q3]. Bold text represents significant between group differences, as assessed by Kruskal-Wallis rank sum test and Fisher's exact test. Definition of abbreviations: WHO FC, World Health Organization Functional Class; FVC, forced vital capacity; FEV1, forced expiratory volume in 1 s; DLCO, diffusion capacity of the lung for carbon monoxide; mPAP, mean pulmonary arterial pressure; AI, artificial intelligence; GGR, ground glass with reticulation

| ROC-AUC | Internal Validation N=17, patches=51,960 | Internal Test N=10, patches=54455 | External Test N=12, patches =41740 |
|---|---|---|---|
| Micro-averaged AUC | 0.92 | 0.95 | 0.94 |
| Macro-averaged AUC | 0.91 | 0.94 | 0.94 |
| Normal classification AUC | 0.96 | 0.96 | 0.97 |
| GG classification AUC | 0.90 | 0.91 | 0.94 |
| GGR classification AUC | 0.87 | 0.88 | 0.87 |
| Honeycombing classification AUC | 0.96 | 0.97 | 0.94 |
| Emphysema classification AUC | 0.87 | 0.99 | 0.96 |

**Table 2: Receiver operator characteristic area under the curve (AUC) for the 5 texture classes with micro and macro averages in the internal validation, internal test and external test cohorts.**

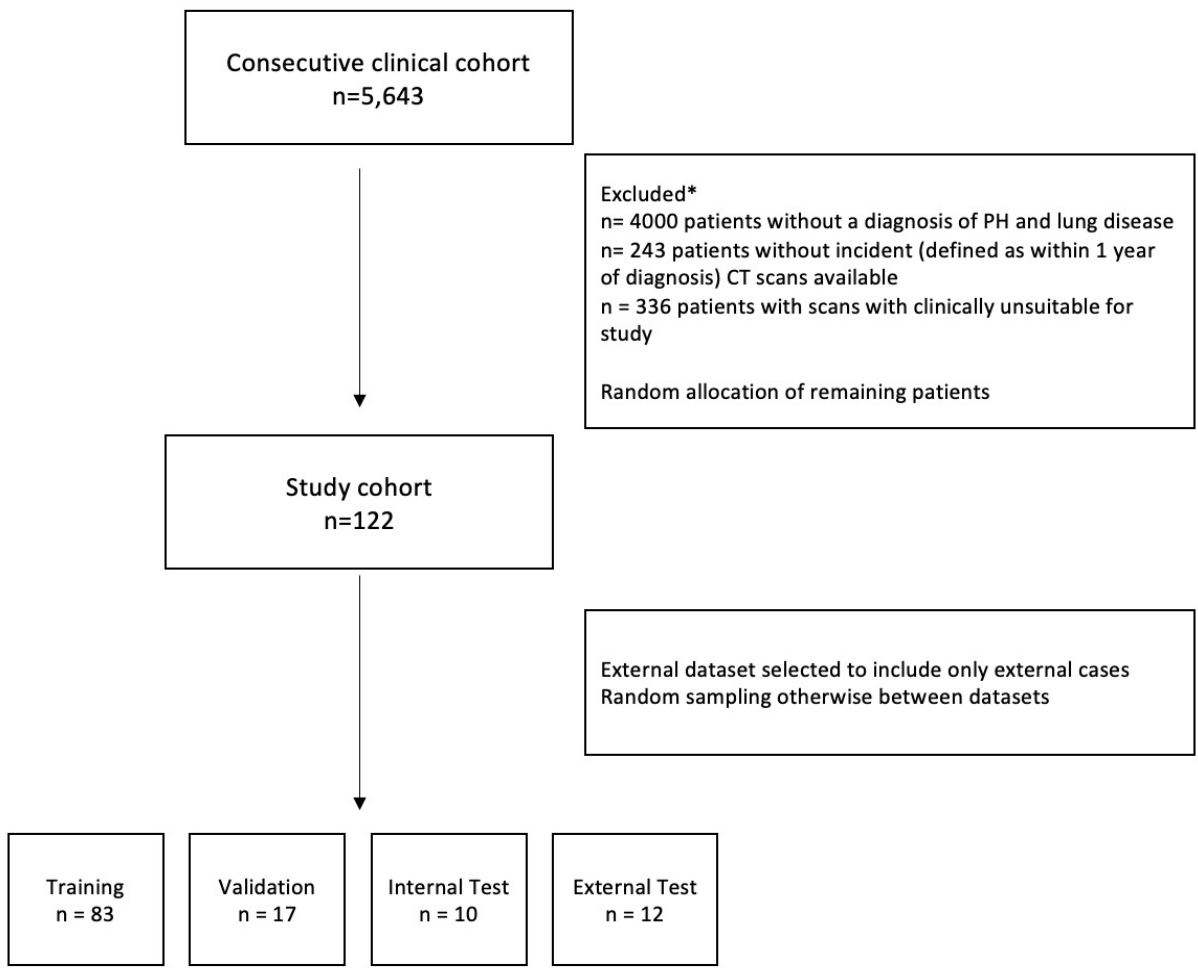

**Figure 1: STROBE diagram showing patient selection and split between datasets for model development and evaluation.** Abbreviation used: CT (Computed Tomography).

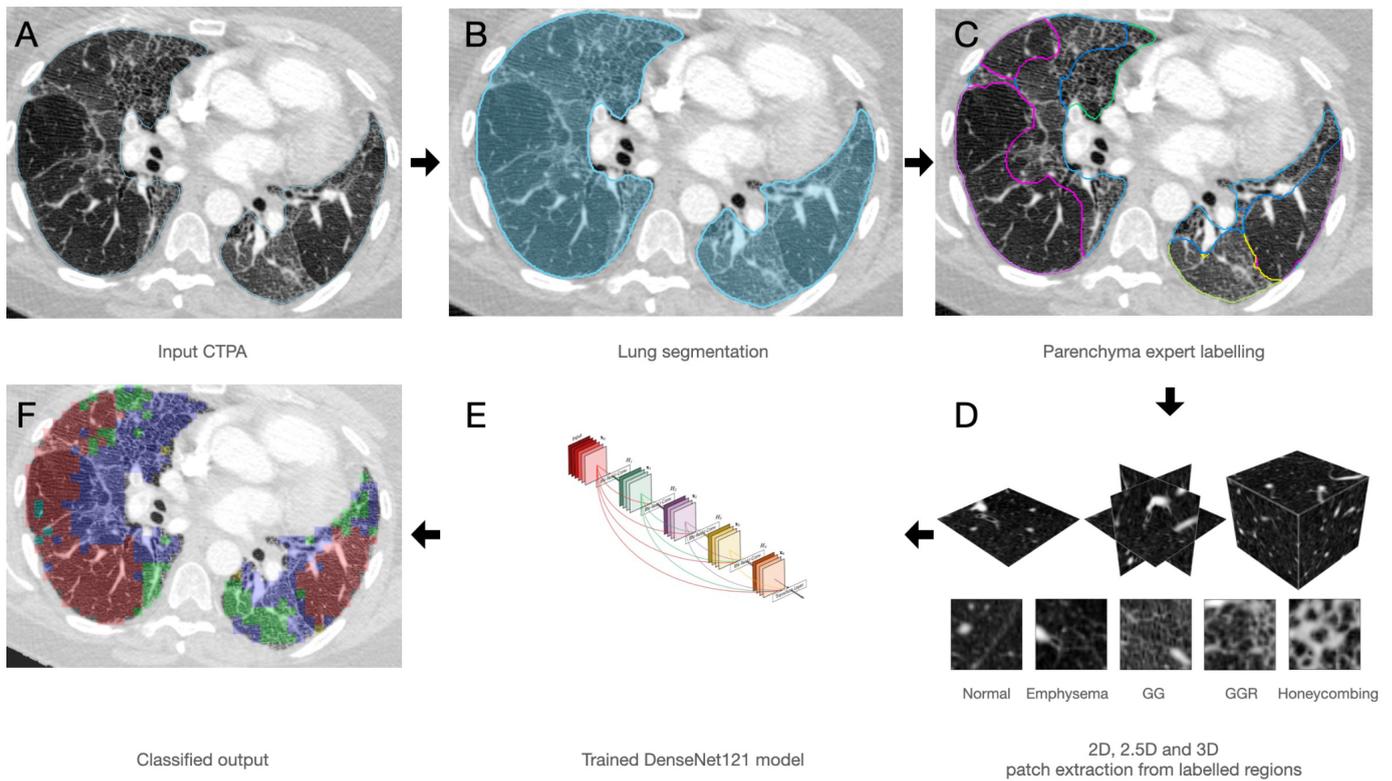

**Figure 2: Schematic of the methodology for patch based approach to classify lung parenchyma.** For each input CTPA (A), lung parenchyma is first segmented (B). Expert radiologists then manually label regions (C) of lung parenchyma which correspond to the five chosen classes of disease - normal, GG, GGR, honeycombing, emphysema and fibrosis. 2D, 2.5D and 3D overlapping patches are created from these labelled regions to form an image atlas (D). The patches are used to train a DenseNet121 model (E). The trained model is subsequently used to classify each voxel of a given input CTPA into one of five classes (F). DenseNet architecture graphic is taken from the original paper describing the method (14).

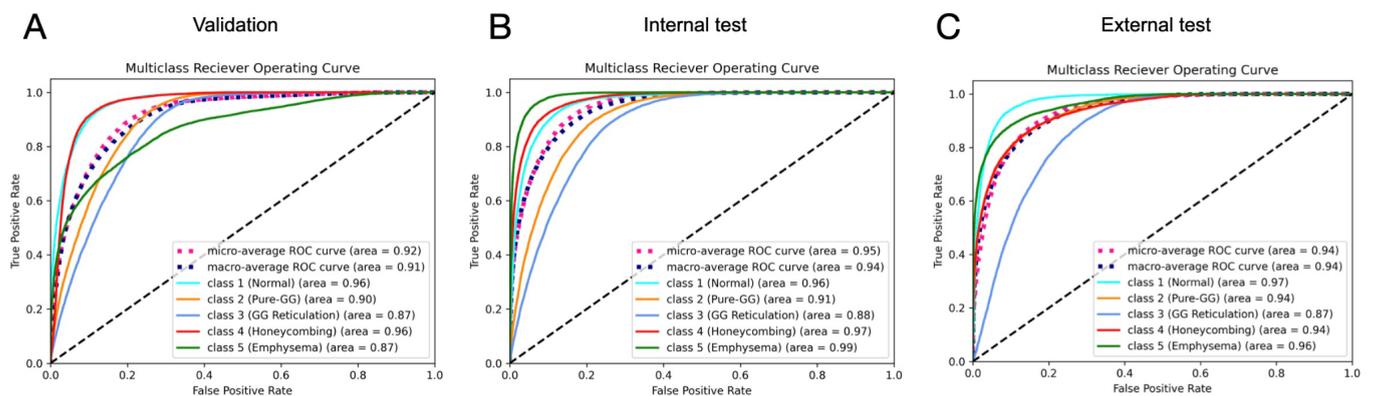

**Figure 3: Multiclass Receiver Operating Curve for validation (A), internal test (B) and external test (C) datasets of the hypertuned DenseNet121 classification model.**

## Author Contributions

Conceptualisation-M.J.S., K.D. and A.J.S.; methodology-M.J.S., K.D., S.A. and A.J.S.; data curation-K.D., M.S., A.J.S.; writing-original draft preparation M.J.S., K.D. and A.J.S.; writing-review and editing; -M.J.S., K.D., S.A. and A.J.S.; supervision; A.J.S.. All authors have read and agreed to the published version of the manuscript.

## Funding

This research is supported by the NIHR Sheffield Biomedical Research Centre. The views expressed are those of the author(s) and not necessarily those of the NIHR or the Department of Health and Social Care. This research was funded in whole, or in part, by the Wellcome Trust [Grant numbers Krit Dwivedi 222930/Z/21/Z and 4ward North 203914/Z/16/; Andrew Swift AJS 205188/Z/16/Z].

## Conflicts of Interest

The Authors declare no conflicts of interest.

## Informed Consent Statement

Research ethics committee approval for retrospective analysis with waiver of informed consent was obtained (ASPIRE, ref: c06/Q2308/8).